\documentclass[aps,prl,preprint,superscriptaddress,tightenlines,nofootinbib]{revtex4}

\usepackage{amsfonts,amsmath,amsthm,amssymb}
\usepackage{mathrsfs}
\usepackage{bm}
\usepackage{color}
\usepackage{epsfig}

\usepackage[usenames,dvipsnames]{xcolor}
\definecolor{orange}{cmyk}{0,0.5,1,0}
\definecolor{rossoCP3}{cmyk}{0,.88,.77,.40}
\definecolor{graa}{rgb}{0.8,0.8,0.8}
\definecolor{blaa}{rgb}{0.2,0.2,0.6}

\newcommand{\PRE}[1]{{#1}}   
\newcommand{\met} {\not\!\! E_T}

\newcommand{\beq}[1]{\begin{equation}\label{#1}}
\newcommand{\eeq}{\end{equation}}
\newcommand{\bea}{\begin{flushleft} \begin{eqnarray}}
\newcommand{\eea}{\end{eqnarray}\end{flushleft}}

\newcommand{\comment}[1]{}

\newcommand{\ci}[1]{}

\newcommand{\ba}{\begin{eqnarray}}
\newcommand{\ea}{\end{eqnarray}}
\newcommand{\be}{\begin{equation}}
\newcommand{\ee}{\end{equation}}
\newcommand{\bay}[1]{\left(\begin{array}{#1}}
\newcommand{\eay}{\end{array}\right)}

\newcommand{\Ecut}{E^\nu_{\max}}
\newcommand{\rarr}{\rightarrow}
\newcommand{\nue}{\nu_{e}}
\newcommand{\numu}{\nu_{\mu}}
\newcommand{\nutau}{\nu_{\tau}}
\newcommand{\nuebar}{\bar{\nu}_{e}}
\newcommand{\nubar}{\bar{\nu}}

\newcommand{\gsim}{\gtrsim}

\def\met{\mbox{${\hbox{$E$\kern-0.6em\lower-.1ex\hbox{/}}}_T$}} 

\newcommand{\beqa}{\begin{eqnarray}}
\newcommand{\eeqa}{\end{eqnarray}}

\newcommand{\la}{\langle}
\newcommand{\ra}{\rangle}

\newcommand{\Eres}{E_{\rm res}}
\newcommand{\Evis}{E_{\rm vis}}

\begin{document}

\title{\PRE{\vspace*{0.9in}} \color{rossoCP3}{ 
    Weinberg's Higgs portal confronting  recent LUX and LHC results 
    together with upper limits on  $\bm{B^+}$ and $\bm{K^+}$ decay into invisibles}
\PRE{\vspace*{0.1in}} }

\title{\PRE{\vspace*{0.9in}} \color{rossoCP3}
{
	End of the cosmic neutrino energy spectrum
}
	
\PRE{\vspace*{0.1in}} }


\author{L.A.~Anchordoqui}

\affiliation{Department of Physics \& Astronomy,\\
Lehman College, City University of New York, Bronx NY 10468, USA
\PRE{\vspace*{.05in}}
}

\affiliation{Department of Physics,\\
University of Wisconsin-Milwaukee,
 Milwaukee, WI 53201, USA
\PRE{\vspace*{.05in}}
}

\author{V.~Barger}
\affiliation{Department of Physics,\\
University of Wisconsin, Madison, WI 53706, USA
\PRE{\vspace*{.1in}}
}

\author{H.~Goldberg}
\affiliation{Department of Physics,\\
Northeastern University, Boston, MA 02115, USA
\PRE{\vspace*{.1in}}
}

\author{J.G.~Learned}
\affiliation{Department of Physics \& Astronomy,\\
University of Hawaii at Manoa, Honolulu, HI 96822, USA
\PRE{\vspace*{.1in}}
}

\author{D.~Marfatia}
\affiliation{Department of Physics \& Astronomy,\\
University of Hawaii at Manoa, Honolulu, HI 96822, USA
\PRE{\vspace*{.1in}}
}

\author{S.~Pakvasa}
\affiliation{Department of Physics \& Astronomy,\\
University of Hawaii at Manoa, Honolulu, HI 96822, USA
\PRE{\vspace*{.1in}}
}

\author{T.C.~Paul}
\affiliation{Department of Physics,\\
University of Wisconsin-Milwaukee,
 Milwaukee, WI 53201, USA
\PRE{\vspace*{.05in}}
}
\affiliation{Department of Physics,\\
Northeastern University, Boston, MA 02115, USA
\PRE{\vspace*{.1in}}
}

\author{T.J.~Weiler}
\affiliation{Department of Physics \& Astronomy,\\
Vanderbilt University, Nashville TN 37235, USA
\PRE{\vspace*{.1in}}
}


\begin{abstract}
 \PRE{\vspace*{.1in}} 
\noindent 

\noindent There may be a high-energy cutoff of neutrino events in
IceCube data.  In particular, IceCube does not observe 
either continuum events above 2~PeV, or
the Standard Model Glashow-resonance events expected at 6.3~PeV.  
There are also no higher energy neutrino signatures in the ANITA and Auger experiments.
This absence of high-energy neutrino events motivates 
a fundamental restriction on neutrino energies above a few PeV.  
We postulate a simple scenario to terminate the neutrino spectrum that is
Lorentz-invariance violating, but 
with a limiting neutrino velocity that is always 
smaller than the speed of light. 
If the limiting velocity of the neutrino applies also to its associated charged lepton,
then a significant consequence is that 
the two-body decay modes of the charged pion are 
forbidden above two times the maximum neutrino energy,
while the radiative decay modes are suppressed at higher energies.
Such stabilized pions may serve as cosmic ray primaries.
 

\end{abstract}
\pacs{xxxx}
\maketitle

%
The fact that IceCube does not (yet) observe neutrinos with energies above
about 2~PeV~\cite{Aartsen:2013bka,Aartsen:2013jdh} invites some
interesting speculation: Perhaps there are none!  
IceCube does observe three events in the 1-2~PeV range.\footnote
{
The two published PeV events deposited
  $1.041^{+0.132}_{-0.144}$~PeV and $1.141^{+0.143}_{-0.133}$~PeV of
  energy in photo-diode electrons (PDEs), respectively.  This PDE
  energy is a minimum but sensible estimate for the true event energy of showers.  
  The third event has the highest
  PDE energy,  $2.004^{+0.236}_{-0.262}$~PeV~\cite{Aartsen:2014gkd}.
  }  
The expected number of continuum (meaning, non-resonant) events above 3~PeV,
normalized to the three observed events in the 1-2~PeV region, 
is 1.5 for a neutrino flux falling as $E^{-2.0}$~\cite{window}.
(Other experiments such as Auger~\cite{Abraham:2007rj} and ANITA~\cite{Gorham:2008yk} are
not necessarily expected to see any neutrinos resulting from Standard
Model (SM) processes, and they do not.)  
Furthermore, the absence of
``Glashow resonance''~\cite{Glashow:1960zz} events ${\bar \nu_e} +e^-
\rightarrow W^- \rightarrow {\rm shower}$ at $E_{\bar \nu_e}=6.3$~PeV
in the IceCube detector volume lends further credence to the cutoff
hypothesis, because the effective area at the resonant energy
partially cancels the falling power law ($E_\nu^{-\alpha}$) of the incident neutrino spectrum.  
The expected number of resonance events (at $\sim 6.3$~PeV) at IceCube for a neutrino flux with flavor ratios 
$\nue:\numu:\nutau=1:1:1$ at Earth as expected from the $\pi^\pm$~production and decay chain,
relative to the three observed events in the 1-2~PeV region is
1.0 for $\alpha=2.0$~\cite{window}.
An earlier statistical study concluded that $\alpha$ was constrained by the
absence of Glashow events in the IceCube data, to $\alpha\ge 2.3$ 
(also assuming a $1:1:1$ flavor ratio at Earth)~\cite{Anchordoqui:2013qsi}.\footnote
{
It is worthwhile to note
  that in $pp$ collisions the nearly isotopically neutral mix of pions
  will create on decay a neutrino population in the ratio
  $N_{\nu_\mu}=N_{\bar\nu_\mu} = 2N_{\nu_e}=2N_{\bar\nu_e}.$ In
  contrast, photopion interactions have the isotopically asymmetric
  process $p\gamma\rightarrow\Delta^+\rightarrow \pi^+ n$,
  $\pi^+\to\mu^+ \nu_\mu\to e^+\nu_e \nubar_\mu \nu_\mu$, as the
  dominant source of neutrinos; then, at production,
  $N_{\nu_\mu}=N_{\bar\nu_\mu} = N_{\nu_e} \gg
  N_{\bar\nu_e}$, 
  which suppresses production of the Glashow resonance~\cite{Anchordoqui:2004eb}. 
  Alternatively, Glashow events can be suppressed if the pion decay chain is terminated in
  the source region by energy loss of the relatively long-lived
  muon~\cite{Kashti:2005qa}.
  }
For a the steeper $\alpha=2.3$ spectrum, 
the expected continuum and resonance event numbers are 1.1 and 0.9~\cite{window}.
The Poisson probability that an experiment would experience a downward fluctuation of 2.5 expected events 
(for $\alpha=2.0$) to zero is $e^{-2.5}=8.2\%$.  
The probability that 2.0 expected events (for $\alpha=2.3$) would fluctuate downward to 
zero is $e^{-2.0}=14\%$. These probabilities are small, but not extremely so.
Our hypothesis is that these numbers are meaningfully small,
and that the absence of events above a cutoff value $\gsim2$~PeV is fundamental.

The showers due to the Glashow resonance (the $W$-boson) 
events are expected to populate
multiple energy peaks~\cite{Kistler:2013my}.  The dominant one is at
$\Eres=M_W^2/2m_e=6.3$~PeV, while the others occur at
$\Evis=\Eres-E_X$, where $E_X$ is the energy in the $W$ decay which
does not contribute to the visible shower: the hadronic decay modes
$W\rarr q\bar{q}$ will populate the peak at 6.3~PeV, while the
leptonic modes $W^-\rarr\nubar+\ell^-$ will lose half of their energy
to the invisible neutrino. Furthermore, the muonic mode will show just
a track, and no shower at all, while in the $\tau$ mode, the $\tau$
decay will produce a second invisible neutrino, leaving a visible
shower with ~1/4 of the energy of the initial $\nuebar$.  Thus we
expect the ratio of the lower energy peaks at 1.6 and 3.2~PeV, to
the higher energy peak at 6.3~PeV, to be
$\rm{BR}(\tau):\rm{BR}(e):\rm{BR(hadrons)}\sim 1:1:6$.  It is tempting to
associate the observed events at 1-2~PeV with the the leptonic decay
modes of the Glashow
resonance~\cite{Barger:2012mz}, but this
exacerbates the issue of why the more numerous resonance events
at 6.3~PeV are not seen.

In this Letter we consider the possibility of a cutoff in the neutrino
energy. The cutoff energy $\Ecut$ could be 2~PeV, 20~PeV, or higher,
with each increase in cutoff energy making the experimental
determination of the cutoff statistically more difficult due to the
expected falloff in the neutrino spectrum (and to some extent, due to
the absence of a null resonance signal for a cutoff energy above
6.3~PeV).  In what follows, we present an example of how a cutoff in
the neutrino energy might arise from Lorentz-Invariance Violation (LIV),
and discuss its phenomenological implications.  If it happens that
there is a bound on the neutrino energy, then
it is possible to discriminate our proposal from standard
cutoff models that do not need new physics ({\it e.g.}, a steeply-falling 
or broken power-law
spectrum~$\propto E_\nu^{-\alpha}$, with $\alpha$
exceeding 2.3). 

The possibility of LIV in terms of a limiting velocity, different for each kind of
particle, has been analyzed in Ref.~\cite{Coleman:1997xq}.
As long as all limiting velocities are less than or equal to
the limiting velocity of the photon, causality is preserved -- new
``lightcones'' appear inside {\it the} lightcone. (For a discussion of
superluminal neutrinos in the present context, see {\it e.g.}~\cite{Gorham:2012qs}.)  
Unlike the realization of LIV of Ref.~\cite{Coleman:1997xq}, in which the limiting velocity 
does not lead to a limiting energy, we postulate an equivalence of limiting energy and limiting velocity $\beta_\nu$:
\beq{Eqone}
\Ecut\equiv \frac{m_\nu}{ \sqrt{ 1-\beta_\nu^2} } \sim
\frac{m_\nu}{\sqrt{2(1-\beta_\nu)}}\,, \quad {\rm with\ } \beta_\nu\equiv \frac{v_\nu}{c}\,.
\eeq
Accordingly, the required limiting velocity to suppress
$\Ecut$~neutrinos is $\beta_\nu\approx 1-1/(2\gamma_\nu^2)$, differing
from the speed of light by 
\begin{equation}
1-\beta_\nu\simeq 0.5\times10^{-28}\left(\frac{m_\nu}{10\,{\rm
      meV}}\right)^2\left(\frac{\rm TeV}{\Ecut}\right)^2\,.
 \end{equation}
%
%
%
It is very likely that the highest boost factors
will forever remain the domain of neutrinos;
even for $\Ecut\ge$~2~PeV, a ``boost-equivalent'' proton
energy is five orders of magnitude higher than has ever been observed
for a proton primary.\footnote {
$E^\nu_{\rm max}$ may be related to a new physics scale, 
such as the size of extra dimensions.
It is
  interesting to note that a proton with boost factor equal
  to that of a PeV~neutrino,
  $\frac{\rm PeV}{m_\nu}\sim 10^{16}$, has an energy of $10^{16}$~GeV,
  comparable to the Grand Unification scale. }
  
The obvious phenomenological statement is that no neutrino with energy
exceeding $\Ecut$ will ever be observed. In particular, so-called
``guaranteed'' BZ~\cite{Beresinsky:1969qj} neutrinos from the
cosmogenic GZK~\cite{Greisen:1966jv} process will not
be produced if $E_{\rm BZ}$, typically $10^{18}$~eV, exceeds $\Ecut$.
However, the $\sim$~PeV $\nuebar$'s from the decay of cosmic ray neutrons
will still be produced.

One may well ask to what reference frame is the limiting velocity
compared?  The Universe has conveniently provided us with a unique
``cosmic rest frame.''  It is the frame where the bulk matter of the
Universe is at rest (nearly the same as the rest frame of Earth).  
Equivalently, it is the frame in which the bulk
temperature of 2.73~K is uniquely and universally defined.  All other
frames are to be compared to this unique Machian reference frame.

Incidentally, although particle limiting velocities will alter the behavior of the very early Universe, 
it is unlikely that there would be any trace of this altered behavior in the later Universe.
This is because the Universe is a thermal system, so alterations would have occured at $T\sim\Ecut \gsim$~PeV, 
long before the QCD transition from quark-gluon-plasma to mesons and baryons 
at $\sim \Lambda_{\rm QCD}\sim 200$~MeV, 
and even longer before nucleosynthesis at $\sim$~MeV.

It seems natural to impose a limiting energy or velocity on each
lepton flavor.  Then in association with the muon (electron) neutrino,
the muon (electron) too has a limiting velocity or energy.
Consequences are significant.  For example, the kinematics of charged
pion decay to leptons and their associated neutrinos having a common
maximum energy $\Ecut$ dictates that charged pions are 
stablized above $\sim2\Ecut$. 
The pion is certainly stable at energies above
$E_{\rm max}^{\nu_\mu} + E_{\rm max}^\mu$ and $E_{\rm max}^{\nu_e} + E_{\rm max}^e$,
if many-body decays are ignored.\footnote{
Unless the unknown mechanism that realizes our postulate suppresses higher-order radiative decays, 
the pion will decay via 
3-body
processes like $\pi^+\to \gamma \ell^+\nu$ and 
$\pi^0 e^+\nue$,
and possibly via 4-body processes like 
$3\ell+\nu$ and {\bf $3\nu+\ell$} (where kinematics allow at most one $\ell$ to be a muon with the others electrons),
and $\pi^0\gamma e^+ \nue$ and $\gamma\gamma\ell\nu_\ell$.
Here we give sufficient energy conditions to ensure $\pi^\pm$~stability; 
in the cases with two final state leptons, the sufficiency condition is calculated with the two leptons 
having identical three-momentum in the direction of the parent $\pi^\pm$.
The necessary conditions on the $\pi^\pm$~energy, not yet available, will be lower than 
the sufficiency conditions given herein.
Sufficient critical $\pi^\pm$~energies for the 3-body modes are
 $\left( {m_{\pi^\pm}^2 \over m_\ell^2}\right) (E_{\rm max}^\nu+E_{\rm max}^\ell)$ and  
 $\left( {m_{\pi^\pm}^2-m_{\pi^0}^2 \over m_e^2}\right) (E_{\rm max}^{\nue}+E_{\rm max}^e)$, respectively
 (with $\ell=\mu$ kinematically forbidden for the final state containing a $\pi^0$).
On the face of it, the sufficient energy assigned to the mode $\pi^+\to \gamma \ell^+\nu$ 
has a dangerous prefactor, $\left( \frac{m_{\pi^\pm}^2}{m_e^2}\right)\sim 0.8\times 10^5$; 
clearly, the lower, necessary energy that bounds this mode needs to be calculated.
The sufficient $\pi^\pm$~energy above which the purely leptonic 4-body decay modes are disallowed are 
$2E_{\rm max}^e +E_{\rm max}^{\ell}+E_{\rm max}^{\nu_\ell}$ and 
$2E_{\rm max}^{\nu_\alpha}+E_{\rm max}^\ell +E_{\rm max}^{\nu_\ell}$, respectively; 
while the sufficient energies to disallow the 4-body semi-leptonic modes,
obtained by setting the lepton pair to rest in the parent $\pi^\pm$-frame,
are the same as the results for the 3-body modes.
We note that since the 3-body modes are special cases of the semi-leptonic 4-body modes 
(with one 4-body photon energy set to zero),
they cannot, and here do not, have a higher requirement on the parent energy.}
%
%
The pion could be stable for all practical purposes at an energy below that dictated by 
absolute stability, because its lifetime may be sufficiently altered by the maximum allowed lepton energies.
Generally speaking, $E_{\rm max}^{\nu_\mu} \neq E_{\rm  max}^\mu \neq E_{\rm max}^{\nu_e} \neq E_{\rm max}^e$,
and so the track to shower ratio may be anomalous as the neutrino energy approaches a PeV.

These absolutely stable high-energy charged pions would
constitute a new primary candidate for cosmic rays with energies above
some stabilization energy;
since pion stability will depend on $E_{\rm max}^\mu$ and $E_{\rm max}^e$ 
as well as on $E_{\rm max}^{\nu_\mu}$ and $E_{\rm max}^{\nu_e} $,
the stabilization energy can be quite different from $E^{\nu}_{\rm max}$.
The possibility of stable pion cosmic ray primaries adds an interesting test of our hypothesis.
The common rigidity (energy/charge) of the protons and charged pions implies that 
(within the Galactic leaky box context) the stable pion spectrum would share the 
slope of the proton spectrum.
On the other hand, the strong Auger limit on the photon content of
ultra high energy cosmic ray showers~\cite{Abraham:2009qb} can be
construed to limit any primary that showers more like a photon than
like a proton.  However, stable pion primaries still maintain their
strong interaction cross section, and so look more like proton
primaries than photon primaries. There are three attributes of a
primary's scattering that determine its typical shower: the cross
section, the inelasticity, and the final state multiplicity.  Stable
pion showers may appear proton-like in the third characteristic, the
final state multiplicity, but they deviate from protons on the first
two attributes: the pion cross section at high energy is smaller by
about 2/3 than the proton cross section, and the pions' Feynman
$x_F$~distribution, a measure of the primary particle elasticity per
interaction, is harder than that of the proton~\cite{Goldberg:1972uf}.
For the pion, $\frac{d\sigma}{dx_F} \propto (1-x_F)$, with a mean
elasticity of $\la x_F \ra=1/3$, while for the proton the cross
section goes as $\propto (1-x_F)^3$, with a mean elasticity $\la x_F
\ra=1/5$.  The proton primary will lose 99\% of its energy after 2.9
interactions on average, whereas the stable pion will lose 99\% of its
energy only after 4.2 interactions.  Thus the shower development is
delayed in the case of a pion primary relative to the proton primary.
A more quantitative study is required to determine whether or not
pion-primaries are viable, useful high-energy cosmic ray
candidates. It is important to note, however, that within this
scenario the stable $\pi^\pm$'s contribute to the total power budget and thereby
reduce the power budget of proton sources 
(e.g., the complete GZK chain reaction proceeds only via $\pi^0$
decay).

In summary, we have explored the hypothesis that there may be an upper limit $\Ecut$ on the neutrino energy.
Such an energy limit implies broken Lorentz invariance, since arbitrarily large boosts
 relative to the cosmic rest frame become
untenable for the energy-bounded neutrino.
IceCube data, statistically weak at present, suggests that this upper limit 
may have already been encountered, at a few~PeV.  

Of course, Ockham's razor favors the absence of the baroque scenario presented here.
The simplest means to raise the search limit for $\Ecut$ 
(and reduce the motivation for our speculation) 
is to observe neutrinos with energies extending to higher and higher values.
However, if the absence of observed neutrinos above some energy persists,
it would be evidence that Nature is more whimsical than William of Ockham.

\section*{Acknowledgments}

LAA is supported by U.S. National Science Foundation (NSF) CAREER
Award PHY1053663 and by the National Aeronautics and Space
Administration (NASA) Grant No. NNX13AH52G. VB is supported by the
U. S. Department of Energy (DoE) Grant No. DE-FG-02- 95ER40896.  HG is
supported by NSF Grant No. PHY-0757959.  JGL is supported by DoE Grant
No. DE-SC0010504.  DM is supported by DoE Grant
No. DE–FG02–13ER42024.  SP is supported by DoE Grant
No. DE-SC0010504 and by the Alexander Von Humboldt Foundation.  TCP is supported by
NSF Grant No. PHY-1205854 and NASA Grant No. NNX13AH52G. TJW is
supported by DoE Grant No. DE-FG05-85ER40226 and the Simons Foundation
Grant No. 306329.


\begin{thebibliography}{99}


\bibitem{Aartsen:2013bka} 
  M.~G.~Aartsen {\it et al.}  [IceCube Collaboration],
  Phys.\ Rev.\ Lett.\  {\bf 111}, 021103 (2013)
  [arXiv:1304.5356 [astro-ph.HE]];

\bibitem{Aartsen:2013jdh} 
 M.~G.~Aartsen {\it et al.}  [IceCube Collaboration],
  Science {\bf 342}, no. 6161, 1242856 (2013)
  [arXiv:1311.5238 [astro-ph.HE]].

\bibitem{Aartsen:2014gkd} 
  M.~G.~Aartsen {\it et al.}  [IceCube Collaboration],
  arXiv:1405.5303 [astro-ph.HE].

\bibitem{window}
V.~Barger, L.~Fu, J.~G.~Learned, D.~Marfatia, S. Pakvasa and T.~J.~Weiler,
arXiv:1407.3255 [astro-ph.HE].


\bibitem{Abraham:2007rj} 
  J.~Abraham {\it et al.}  [Pierre Auger Collaboration],
  Phys.\ Rev.\ Lett.\  {\bf 100}, 211101 (2008)
  [arXiv:0712.1909 [astro-ph]];
  Phys.\ Rev.\ D {\bf 79}, 102001 (2009)
  [arXiv:0903.3385 [astro-ph.HE]];
  P.~Abreu {\it et al.},
  Phys.\ Rev.\ D {\bf 84}, 122005 (2011)
  [arXiv:1202.1493 [astro-ph.HE]];
  Astrophys.\ J.\  {\bf 755}, L4 (2012)
  [arXiv:1210.3143 [astro-ph.HE]].



\bibitem{Gorham:2008yk} 
  P.~W.~Gorham {\it et al.}  [ANITA Collaboration],
  Phys.\ Rev.\ Lett.\  {\bf 103}, 051103 (2009)
  [arXiv:0812.2715 [astro-ph]];
  Phys. Rev. D {\bf 82} 022004 (2010)
 [arXiv:1003.2961 [astro-ph.HE]];
  Phys.\ Rev.\ D {\bf 85}, 049901 (2012)
  [arXiv:1011.5004 [astro-ph.HE]].





\bibitem{Glashow:1960zz} 
  S.~L.~Glashow,
  Phys.\ Rev.\  {\bf 118}, 316 (1960).














\bibitem{Anchordoqui:2013qsi} 
  L.~A.~Anchordoqui, H.~Goldberg, M.~H.~Lynch, A.~V.~Olinto, T.~C.~Paul and T.~J.~Weiler,
  [Phys. Rev. D, in press arXiv:1306.5021 [astro-ph.HE]];
  L.~A.~Anchordoqui, V.~Barger, I.~Cholis, H.~Goldberg, D.~Hooper, A.~Kusenko, J.~G.~Learned, D.~Marfatia, S. Pakvasa, T. C. Paul and T. J. Weiler,
  JHEAp {\bf 1}, 1 (2014)
  [arXiv:1312.6587 [astro-ph.HE]].

\bibitem{Anchordoqui:2004eb} 
  L.~A.~Anchordoqui, H.~Goldberg, F.~Halzen and T.~J.~Weiler,
  Phys.\ Lett.\ B {\bf 621}, 18 (2005)
  [hep-ph/0410003].



\bibitem{Kashti:2005qa} 
  T.~Kashti and E.~Waxman,
  Phys.\ Rev.\ Lett.\  {\bf 95}, 181101 (2005)
  [astro-ph/0507599].


\bibitem{Kistler:2013my} 
  M.~D.~Kistler, T.~Stanev and H.~Yuksel,
  arXiv:1301.1703 [astro-ph.HE].



\bibitem{Barger:2012mz}
  A.~Bhattacharya, R.~Gandhi, W.~Rodejohann and A.~Watanabe,
  JCAP {\bf 1110}, 017 (2011)
  [arXiv:1108.3163 [astro-ph.HE]];
  V.~Barger, J.~Learned and S.~Pakvasa,
  arXiv:1207.4571 [astro-ph.HE];
  A.~Bhattacharya, R.~Gandhi, W.~Rodejohann and A.~Watanabe,
  arXiv:1209.2422 [hep-ph].


  

\bibitem{Coleman:1997xq} 
  S.~R.~Coleman and S.~L.~Glashow,
Phys.\ Lett.\ B {\bf 405}, 249 (1997)
[hep-ph/9703240];
Phys.\ Rev.\ D {\bf 59}, 116008 (1999)
[hep-ph/9812418].

\bibitem{Gorham:2012qs} 
  P.~W.~Gorham {\it et al.},
  Phys.\ Rev.\ D {\bf 86}, 103006 (2012)
  [arXiv:1207.6425 [astro-ph.HE]];
  F.~W.~Stecker,
  Astropart.\ Phys.\  {\bf 56}, 16 (2014)
  [arXiv:1306.6095 [hep-ph]].


\bibitem{Beresinsky:1969qj} 
  V.~S.~Berezinsky and G.~T.~Zatsepin,
  Phys.\ Lett.\ B {\bf 28}, 423 (1969).


\bibitem{Greisen:1966jv} 
  K.~Greisen,
  Phys.\ Rev.\ Lett.\  {\bf 16}, 748 (1966);
  G.~T.~Zatsepin and V.~A.~Kuzmin,
  JETP Lett.\  {\bf 4}, 78 (1966)
  [Pisma Zh.\ Eksp.\ Teor.\ Fiz.\  {\bf 4}, 114 (1966)].


    

\bibitem{Abraham:2009qb} 
  J.~Abraham {\it et al.}  [Pierre Auger Collaboration],
Astropart.\ Phys.\  {\bf 31}, 399 (2009)
[arXiv:0903.1127 [astro-ph.HE]];
  Astropart.\ Phys.\  {\bf 29}, 243 (2008)
  [arXiv:0712.1147 [astro-ph]].




\bibitem{Goldberg:1972uf} 
  H.~Goldberg,
  Nucl.\ Phys.\ B {\bf 44}, 149 (1972).


\end{thebibliography}
\end{document}